\documentclass[epj]{svjour}
\usepackage[numbers,sort&compress]{natbib}

\usepackage{epsfig,graphicx,tabularx,epstopdf,float}

\begin{document}
\title{Collisional dynamics of  multiple dark solitons in a toroidal 
Bose-Einstein condensate: Quasiparticle picture}
\author{ H. M. Cataldo\inst{1,2} \and D. M. Jezek\inst{1,2}}
\institute{Universidad de Buenos Aires. Facultad de Ciencias Exactas y Naturales.
Departamento de F\'{i}sica. Buenos Aires, Argentina. \and
IFIBA, CONICET, Pabell\'on 1, Ciudad Universitaria, 1428 Buenos Aires, Argentina.}

\date{Received: date / Revised version: date}

\abstract{
We study the collisional dynamics of multiple dark solitons in a Bose-Einstein condensate 
confined by a  toroidal trap.
We assume a tight enough confinement in the radial direction to prevent possible dissipative effects due to the
presence of solitonic vortices. Analytical expressions for the initial order parameters with imprinted phases
 are utilized to generate different initial arrays of solitons, for which
the time-dependent  Gross-Pitaevskii equation is numerically solved.
Given that the soliton velocity is conserved due to the lack of dissipation,
 we are able to apply a simple quasiparticle  description  of the soliton dynamics. In fact, the  trajectory
equations are written
 in terms of the  velocities 
and the angular shifts
produced at each collision, in analogy to the infinite one-dimensional system. 
% In a first step, we establish for this scenario a relationship between shifts 
%by assuming the conservation of the soliton angular momentum.
To calculate the angular shifts, we directly extract them
 from the trajectories  given by the Gross-Pitaevskii simulations
and, on the other hand, we show that  accurate values can be analytically obtained  by adapting a
formula valid for the  infinite one-dimensional system that involves the healing length, which in
our inhomogeneous system
 must be evaluated  in terms of the sound velocity along the azimuthal direction. 
We further  show that  very good estimates of such a sound velocity can be
directly determined by using the ground state density profile  and the values of the imprinted phases.
We discuss the possible implementation of the system here proposed using
the current experimental techniques.
}
\authorrunning{H. M. Cataldo and D. M. Jezek}
\titlerunning{Collisional dynamics of  multiple dark solitons}
\maketitle
\section{Introduction}\label{sec1}

Solitons arise as fundamental solutions of nonlinear wave equations ruling in quite diverse systems  such as, 
shallow liquid waves \cite{denardo}, magnetic films \cite{chen}, complex plasmas \cite{heide}, 
and optical fibers \cite{kivshar}. 
Particularly, in Bose-Einstein condensates (BECs)
solitons are characterized by their form stability under time evolution, 
even after interacting with other solitons, 
behaving akin to classical particles. 
According to the nature of the interatomic interaction occurring in such ultracold gases, 
that is, an attractive or repulsive one, we may respectively have bright or dark solitons, of which the latter
will be in the focus of the present work.

 Soliton collisions have been extensively studied in infinite one-dimensional (1D) systems
 from the theoretical viewpoint \cite{tsu71,zak73,libro1,libro,theo10}. 
 In such systems, solitons collide elastically and continue moving with a constant velocity away from 
 the collision region. Hence,
the dynamics of a number of interacting solitons can be described by considering them as 
quasiparticles moving with constant velocities, whereas the corresponding collisions are 
regarded as instantaneous
shifts in the soliton positions.
In a pioneering work, Tsuzuki arrived at a simple relationship between the shifts
produced at a collision of two solitary waves \cite{tsu71} 
and later, in Ref. \cite{zak73},
an explicit expression for the values of such shifts was obtained. In this formalism each soliton is identified 
by the corresponding velocity.
In a more recent work, it has been shown that  very  slow solitons, which are identified by their density notches,  
can be safely  regarded  as 
a hard-sphere-like system of
particles, which interact through an
effective (velocity dependent) repulsive potential \cite{theo10}. 

As strictly 1D condensates are impossible to implement experimentally, 
 more realistic  soliton systems in  atomic BECs confined by different trapping potential geometries 
 have been extensively studied  in the last years 
\cite{ rev10}. Recently, renewed interest has arisen from the experimental observation of solitonic vortices in
bosonic and  fermionic systems \cite{donadello14,ku14,chevy14}. In such  BEC 
experiments, solitons have been spontaneously created  through the Kibble-Zurek mechanism \cite{lamporesi13}.
The so far commonly used candidate  to experimentally realize a realistic configuration close to that of the 
infinite 1D system 
 has been a cigar-shaped condensate. Such a condensate, however, presents the potential drawback of showing 
a quite different 
(oscillating) single soliton dynamics. It has been shown that in the Thomas-Fermi approximation a soliton oscillates in a cigar-shaped condensate
with the frequency $  \omega_{trap} / \sqrt{2}$ \cite{theo10,fedi99,bus00,konotop04,becker08,mass1}, where $\omega_{trap}$
 is the angular 
frequency
of the trap in the longitudinal direction.
One could avoid, however,  such a potentially undesirable effect 
stemming from the harmonic trap, simply
 by changing to a toroidal condensate \cite{je16,gal16} . 

 Experimental setups of toroidal condensates have been extensively  utilized for 
different purposes \cite{boshier13,jen14,toroidal}, particularly, they
are currently employed as a fundamental component of test-bed configurations  
within the emerging field of atomtronics  \cite{seaman,boshier13,jen14,eckel}.
On the other hand, the experimental realization of solitonic initial profiles in  such type of condensates
could in principle be implemented by 
standard phase imprinting 
and/or density  manipulation methods, similarly to those applied in cigar-shaped condensates
\cite{becker08}.  Actually, an improved phase imprinting protocol 
for preparing states of given circulation in a toroidal condensate has been recently proposed
\cite{perrin}. In such work, the authors stated that it would be very interesting to extend these
ideas to create multiple solitons with well defined relative velocities.
In a recent work \cite{je16}, we have derived an expression for the soliton energy in a toroidal configuration,
which depends on the imprinted phases. In particular, we have shown that
such an energy turns out to be a decreasing function of the soliton velocity.
In such a work, we have studied the dynamics of a pair of symmetrically
 counter-rotating solitons in a toroidal
condensate for a wide range of initial velocities. We note that the soliton velocity 
$v$ lies between $0$ and $c$ ($ 0 < v < c$), where $c$ is the sound velocity. 
It has been shown that only for very slow solitons ($ v \simeq  0.001 c$),
the  angular velocity remains constant along the evolution except, of course, around the collisions. 
For larger velocity values, the solitonic profiles are converted in solitonic vortices. 
In such a work, we were
interested in describing the active role that the vortices play in the dynamics. As a consequence of the appearance
of solitonic vortices, a continuous increase of the soliton velocities along the evolution was observed,
which
in turn yields a decrease of the soliton energy that can be interpreted as a dissipative dynamics of the solitonic system. 
Several other regimes arise from a modulation of the trap parameters  \cite{gal16}.

The goal of this work is to study the non dissipative dynamics  of gray solitons 
 in a toroidal BEC using a simple quasiparticle picture. We will show that the trajectories,  
  described  in terms of the soliton velocities and the angular shifts, can be determined by solely using 
the density maximum of a 
Gaussian ground state profile and the imprinted phases.
%which includes asymmetric  collisions  and 
 For that purpose, we will model a toroidal condensate  tightly enough  confined in the radial direction,  
 in order to discard sources of dissipation associated to the presence of solitonic vortices, which
may affect the conservation of the soliton velocities.
For such trapping parameters, we have found that energy dissipation occurs 
as the soliton velocity increases during the time evolution. In the present work 
we will analyze the soliton trajectories
arising from time-dependent Gross-Pitaevskii (GP) simulations.
In particular, we will demonstrate that
the soliton velocities remain constant along the evolution and we will calculate the angular shifts at the collisions. 
We will show that such shifts can be accurately calculated
by only precisely determining the sound velocity for each set of imprinted phases.
 
It is important to notice that the soliton velocity determines the size of the 
condensate density notch and hence the ``negative" mass
of the soliton. Thus, the quasiparticle picture involves solitons with a definite mass.
We will  show that  a rich variety of
initial configurations may be reached by implementing certain phase-imprinting protocol.

This paper is organized as follows.
In Sec. \ref{sec2} we introduce the system, particularly the toroidal trap and the set of parameters involved, 
which are chosen in order
to avoid sources of dissipation.
In Sec. \ref{sec3},  based on a previous work \cite{je16},
 we propose a form of constructing initial arrays of  gray solitons with different imprinted velocities.
Section \ref{sec4} is devoted to the study of the soliton dynamics. We first obtain
 the soliton trajectories along the torus by solving the time-dependent 
GP equation, showing that in fact, the velocities are conserved, and hence, a simple quasiparticle picture 
for describing the dynamics can be applied.
Within such a picture the quasiparticle  collisions are described as angular shifts.
A relationship  between the angular
 shifts involved in a collision, analogous to that of the 1D system, is established
and, on the other hand, such shifts are evaluated using the trajectories obtained from GP simulations.
We also  show that very accurate shift values can be analytically obtained by using a few parameters,
namely the maximum of the ground state density and the imprinted phases.
In addition, we show that the velocity of sound propagating along the angular direction acquires a 
relevant role 
in determining  the accuracy of the model, 
and thus we analyze the dependence of such velocity on the imprinted phases. 
Finally, the 
conclusions of our study are gathered in Sec. \ref{sec5}.
 
\section{Theoretical framework }\label{sec2} 

We assume a toroidal
 trapping potential written as the sum of a term depending only
on $x$ and $y$, and a term that is harmonic in the $z$ direction:
\begin{equation}
V_{\rm{Trap}}(x,y,z)= V(r) + \frac{1}{2} m  \Omega_z^2  z^2 \, ,
\end{equation}
where  $r^2=x^2+y^2$ and  $m$ denotes the atomic mass of  $^{87}$Rb. 

The term depending on $r$ is modeled as the following  ring-Gaussian potential 
\cite{mu13}
\begin{equation}
V(r)=    V_0 \, \left[  1 - \exp\left[ - \Lambda \left(\frac{r}{r_0}-1 \right)^2\right]   \right],
\label{toro}
\end{equation}
where $V_0$ and  $r_0$   denote the depth and  radius 
of the potential minimum. The dimensionless parameter $\Lambda $ is associated to the $1/e^2$
potential width
$ w = r_0 \sqrt{ \frac{ 2}{\Lambda} }$.

The trap parameters have been selected to reproduce similar 
experimental conditions to those described in Ref.
\cite{boshier13}. We have set  $V_0$=70 nK, $r_0$=4 $\mu \rm{m} $, and fixed the particle number to
$ N=1000$.
A high value of $\Omega_z=2\pi\times 922$ Hz  yields
a quasi two-dimensional (2D) condensate which allows a simplified numerical treatment \cite{castin}.
Then, the order parameter can be represented as a product
of a wave function on the $x$-$y$ plane, $\psi(x,y)$,
and a Gaussian wave function along the $z$ coordinate,  from which
the following 2D interacting  parameter can be extracted  \cite{castin}
\begin{equation}
g=g_{3D}\left(\frac{m\Omega_z}{2\pi\hbar}\right)^{1/2},
\end{equation}
where $g_{3D}=4\pi\hbar^2a/m$,
being $a= 98.98\, a_0 $ the  
$s$-wave scattering length of $^{87}$Rb and $a_0 $ being the Bohr radius.

In the mean-field approximation, the condensate dynamics is ruled by the time-dependent
 GP equation %\cite{gros61,*pita}
\begin{equation}
i\hbar\frac{\partial \psi}{\partial t}=\left[-\frac{ \hbar^2 }{2 m}{\bf \nabla}^2  +
V(r)+g\, |\psi|^2\right]\psi,
\label{gp}
\end{equation}
where $ \psi\equiv\psi(x,y,t)$ denotes a 2D
order parameter normalized to the number of particles.
Finally, as will be discussed in the next section, 
a high  value of the dimensionless parameter $\Lambda=50$ ($ w =  0.8 $ $\mu \rm{m}$)
is assumed,  in order to assure a 
 condensate confinement in the radial direction 
that avoids the appearance of solitonic vortices in the dynamics.

\section{Initial arrays of gray solitons }\label{sec3}
In a previous work \cite{je16},  we have shown that for a similar toroidal trap,
a dark soliton located along the $x$-axis may be safely modeled through
 an order parameter with a  Gaussian profile of the form,
\begin{eqnarray}
\psi_{G}(r,\theta) &=& \sqrt{n(X)} \, \exp \left[ - \frac{\gamma}{2}\left(1-\frac{r}{r_0}\right)^2
\right] \nonumber\\
 & \times &  \left[ \sqrt{1- 
X^2}  \tanh\left(\sqrt{1- X^2}\,     {\rm k} r \sin\theta\right) + i X \right], \nonumber\\
\label{gaus}
\end{eqnarray}
with $  {\rm k} = \sqrt{n(X=0) g m}/ \hbar$  and   $  n(X)=  n_0/ ( 1 - \frac{2 \sqrt{1- X^2}}{ \pi   {\rm k}   r_0})$,
where   $0\le X\le 1$, and $n_0$ denotes the ground-state density  maximum located at $ r= r_0 $. 
For $X =1$ one recovers the ground state, whereas for $X =0$  a stationary state with 
a double-notch with vanishing density  is obtained,  which is  referred to as a black soliton.
Gray solitons are determined  by  the intermediate values $0< X <1 $,  which  define the absolute value of the
soliton initial velocity  in units of the sound speed.  Such gray solitons are characterized by 
 having  non vanishing density notches.
It is important to note that the subsequent dynamics of the  gray soliton  initial order parameter
actually corresponds to a pair of symmetrically counter-rotating solitons, 
with the soliton initially located across the torus at $x<0$
($x>0$) moving clockwise (counterclockwise) \cite{je16}. As the number of particles is fixed, $  n(X)$ is
a decreasing function of $X$, which is due to the fact that for a smaller $X$ more particles are 
expelled from the density notch.
It is easy to propose a generalization of the above ansatz for an initial state consisting
 of an even number $N_S$ 
of rotating solitons:
\begin{eqnarray}
&\psi_{G}(r,\theta) &=  A \, \exp \left[ - \frac{\gamma}{2}\left(1-\frac{r}{r_0}\right)^2 
\right] \nonumber\\
&&\prod_{j=1}^{N_S/2}  \left[ \sqrt{1- 
X_j^2}  \tanh\left(\sqrt{1- X_j^2}\,   {\rm k}  r \sin (\theta-\alpha_j) \right) \right. \nonumber\\
&&+ i X_j \left.\right] ,  
\label{2gaus}
\end{eqnarray}
being $ A $   a normalization constant 
 and   $   {\rm k} = \frac{\sqrt{ n_0 g m}}{ \hbar}$.
Here the pair of gray solitons labeled by the subscript $j$ are assumed to be initially located along an axis
forming an angle $ \alpha_j$ with the $x$-axis, and they move counter-rotationally at an angular speed determined
by the parameter $X_j$ ($0 < X_j < 1$). 
We note that the phase between adjacent density notches turns out to be a flat function of $\theta $,
i.e., it does not present any gradient around the torus, which would
be the case if single-notch solitons were generated for each  parameter $X_j$ 
\cite{perrin}.  

\begin{figure*}
\centering
\includegraphics{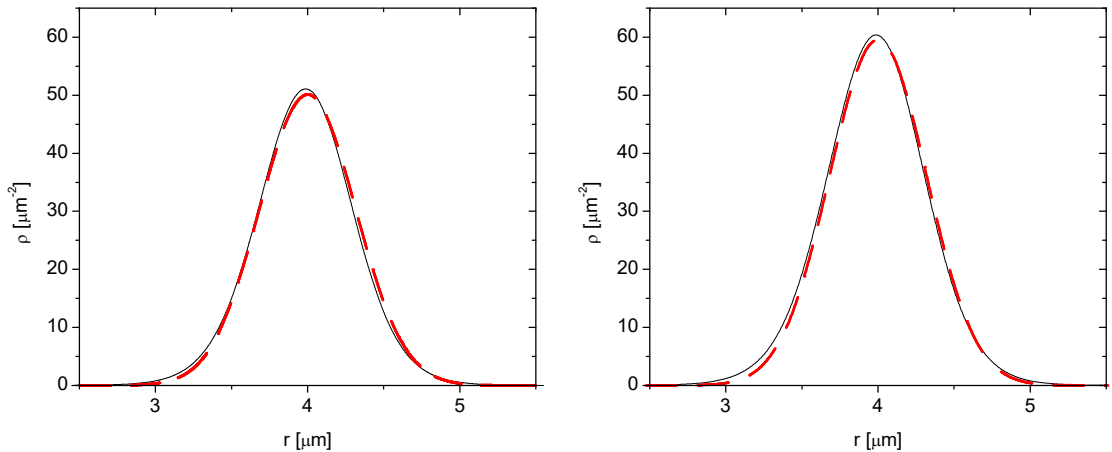}
\caption{Density profiles 
$\rho(r,\theta)=|\psi(r,\theta)|^2$ as functions of the radial coordinate $r$ for the ground state (left panel),  
and the  four-notch black soliton (right panel) calculated  at $\theta=\pi/4$. 
Black solid lines correspond to the GP density, whereas
the red dashed lines correspond to the Gaussian model (\ref{2gaus}) 
for   $X_j=1$ (left panel) and $X_j=0$ (right panel) with  $\alpha_1=0$ and $\alpha_2= \pi/2 $. 
 }
\label{figu1}
\end{figure*}

\begin{figure*}
\centering
\includegraphics{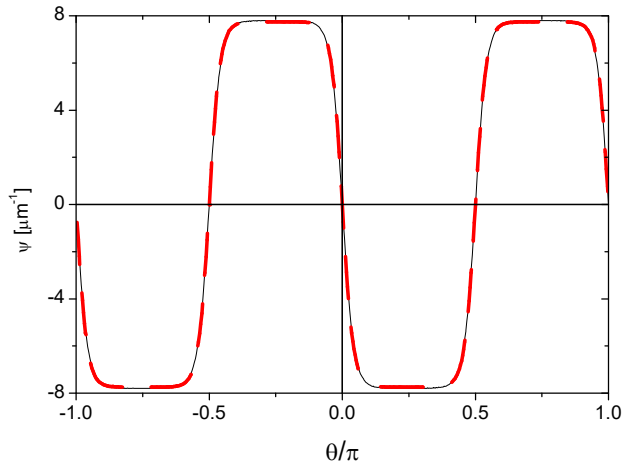}
\caption{The four-notch black soliton order parameter as a function of the angular coordinate calculated at
the potential minimum $r=r_0=4\,\mu$m.  
The black solid line corresponds to the GP solution, whereas
the red dashed line corresponds to the Gaussian model (\ref{2gaus}) 
for  $X_j=0$  with  $\alpha_1=0$ and $\alpha_2=  \pi/2 $. }
\label{figu2}
\end{figure*}

In Fig.~\ref{figu1}, we depict the densities corresponding to the stationary solutions of the GP equation
for the ground state ($X_j=1$)
and the four-notch black soliton ($X_j=0$), which are very similar to those
 predicted by the Gaussian model (\ref{2gaus}) with $ {\rm k} =1.95\, \mu$m$^{-1}$. 
It is worthwhile noticing that the density maximum of the four-notch black soliton configuration
turns out to be appreciably higher than that of the
ground state, because a large number of atoms are expelled from the density notches.
Fig. \ref{figu2} shows 
the GP order parameter of the four-notch black soliton as a function of the angular coordinate for $r=r_0$,
which shows a very good agreement with that given by the Gaussian model, as well.
Such a profile also quantitatively agrees with the kinks observed in the second-excited
stationary analytic solution of a strictly 1D ring ruled by the nonlinear Schr\"odinger equation \cite{carr},
and the same agreement was observed between the double-notch black soliton and the two-node first-excited solution of the 1D ring system \cite{je16}.

\begin{figure*}
\includegraphics{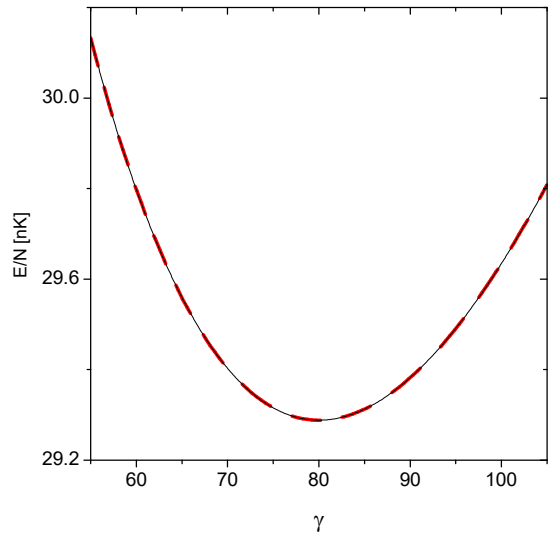}
\centering
\caption{ Energy per particle corresponding to the soliton-free order parameter (\ref{2gaus}) 
as a function of the Gaussian parameter $\gamma$. 
The red dashed  line corresponds to the analytic expression given by 
  Eq. (\ref{enervar}), whereas the  black solid  line corresponds
to a numerical integration of the energy.  }
\label{figu3}
\end{figure*}

The value of $ \gamma $ in Eq.~(\ref{2gaus})
has been determined by minimizing the energy per particle,
\begin{eqnarray}
E( \gamma) & = & \frac{1}{N}\int d^2 {\bf r}\,\,  \psi_G(r, \theta, \gamma)   \left[-\frac{ \hbar^2 }{2 m}{\bf \nabla}^2  +
V(r)\right. \nonumber \\
& + &  \left.\frac{g}{2}\, | \psi_G(r, \theta, \gamma)|^2\right]  \psi_G(r, \theta, \gamma)  ,
\label{ene}
\end{eqnarray}
where $ \psi_G(r, \theta,\gamma) $  is  the
order parameter given by Eq. (\ref{2gaus}) with $X_j=1$ and  $\gamma$ is used as a variational parameter.
We depict in Fig.~\ref{figu3} such an energy as a function of $\gamma$, 
whose minimum turns out to be around 80.
On the other hand, we have found 
the following analytical approximation  valid for  $\gamma>>1$,
\begin{equation}
E( \gamma)   =  \frac{\hbar^2}{4 M r_0^2}  \, \gamma  +   V_0   \left[  1 - \sqrt{\frac{\gamma}{\Lambda + \gamma }} \right]
+ \frac{ N g}{2^{\frac{5}{2}} \pi^{3/2} r_0^2} \, \gamma^{1/2},
\label{enervar}
\end{equation}
where the first, second and third term on the right-hand side, correspond to
the kinetic, trap and interaction energies, respectively. 
Such an expression shows an excellent agreement with
the numerically integrated energy, as seen in Fig.~\ref{figu3}.
We have found that the error in using such an expression turns out to be less than
0.01 \% for $\gamma>30$. Therefore, the value of $\gamma$ for the energy minimum can be safely
obtained by using Eq. (\ref{enervar}) for our parameter range, which in fact yields $\gamma=80$.

The values of the trap parameter $\Lambda$ and the number of 
particles  have been selected according to a recent work by 
 Gallucci and Proukakis \cite{gal16},
where the authors studied the regimes that appear in the dynamics
when varying 
the dimensionless parameter $ l_r /\zeta_0 $  for a similar trapping potential,
where $l_r= \sqrt{\hbar/m \omega_r}$ corresponds to the harmonic oscillator length and 
 $ \zeta_0  = \frac{\hbar}{\sqrt{ n_0 g m}} $ denotes the healing length.
 Three distinct regimes were identified: solitonic (stable), shedding, and snaking (unstable).
In particular, the solitonic regime features an internal subdivision around $ l_r /\zeta_0 =1 $ \cite{gal16}.
 Here we want to avoid  the appearance of solitonic vortices 
and thus we  restrict ourselves to the  regime defined by $ l_r /\zeta_0 < 1$, where radial excitations are
 suppressed. We have
found that with a high $\Lambda=50$ and a number of particles $N=1000$ such a relation is verified.
 More precisely,
the size of the condensate in the radial direction may be directly estimated  from the expression 
 of the Gaussian density compared to that of a 
harmonic oscillator trapping potential:
\begin{equation}
\psi^2_{G}(r)=  A \, \exp \left[ - \gamma\left(1-\frac{r}{r_0}\right)^2\right] \, = 
A \, \exp \left[ - \frac{\left(r- r_0\right)^2}{l_r^2}\right] 
\label{2gausgs}
\end{equation}
leading to $ l_r= r_0/\sqrt{\gamma} =  0.447 $ $ \mu$m.  On the other hand, we have
$ \zeta_0=  0.51  $ $\mu$m for the healing length,  which confirms us that we
are within the pursued regime.
Here it is worthwhile noticing that in our previous work  \cite{je16}, these quantities yielded 
 $ l_r=   0.685$ $\mu$m and  $ \zeta_0=  0.50$ $\mu$m, consistent with the fact that
we were also interested in exploring how the presence of  vortices affects the dynamics.
In fact, in such a work it was shown that the solitons become accelerated, 
a signature of dissipative effects that appear when solitonic vortices are formed.

\section{The dynamics}\label{sec4}

\subsection{ GP numerical simulations \label{GPsim}}

\begin{figure*}
\centering
\includegraphics{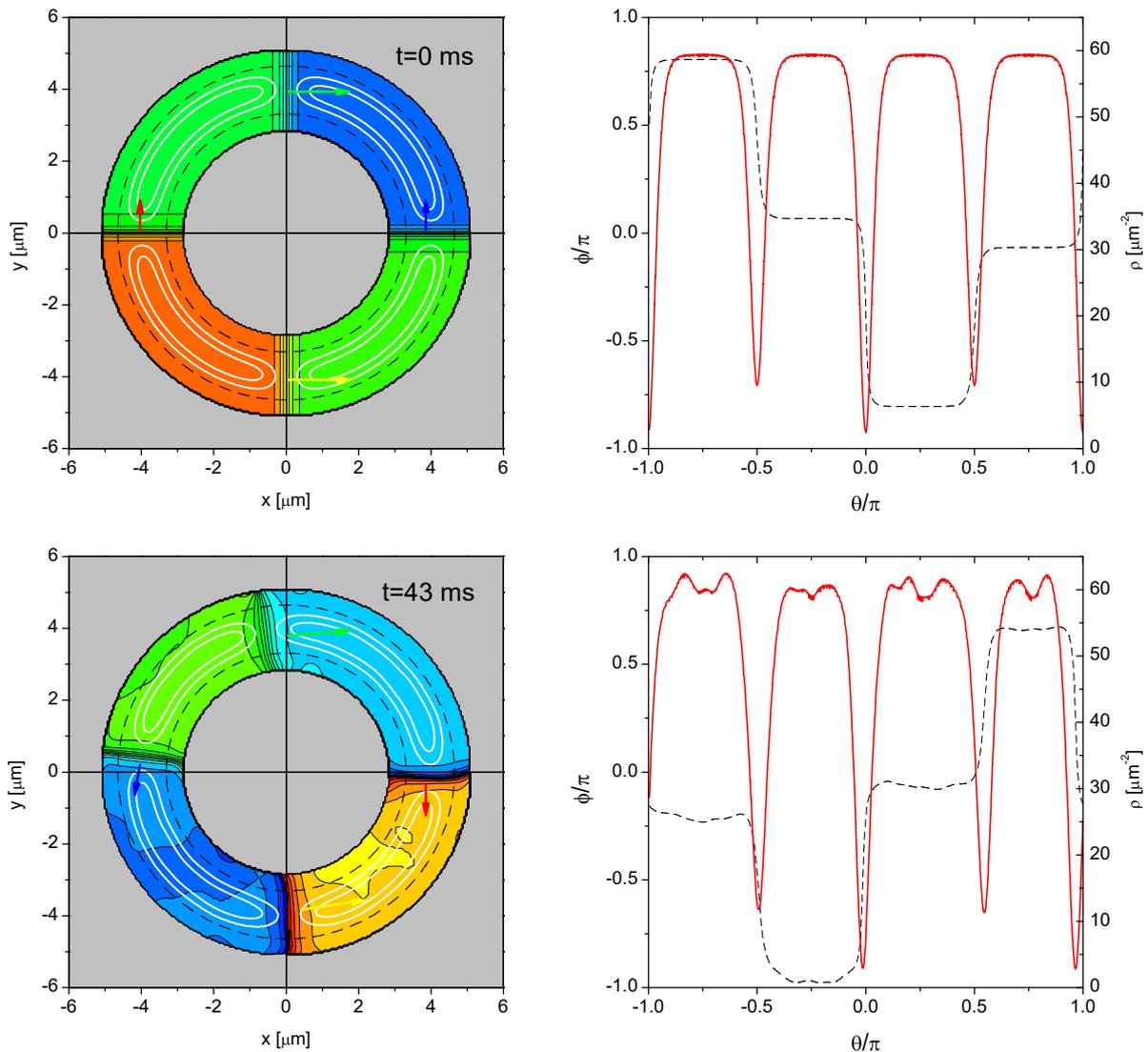}
\caption{
In the left panels we show the phase distribution (colors) and particle density isocontours 
(white solid lines) for a GP time dependent simulation with an initial condition obtained from Eq. (\ref{2gaus}) 
for $X_1=0.2$ and $X_2=0.4$. 
The top and bottom panels correspond to snapshots taken at $t=0$ and  $t=43$ ms, respectively.
A dashed line is also shown at the left panels to denote isocontours corresponding to a
10 percent of the maximum density of the ground state, while the arrows indicate the soliton velocities
and their colors identify each soliton.
The right panels
show angular distributions of the particle density (red solid line)  and phase 
(black dashed line) for
the azimuthal angle $ \theta $ at the radius $r_0= 4 \, \mu$m. The phase colors at the left panels are 
linked to
the corresponding phase values depicted on the right panels. }
\label{figua}
\end{figure*}

To analyze the soliton dynamics, we have numerically  solved the time-dependent GP equation
for initial order parameters of the form (\ref{2gaus}) with different soliton velocities.
For all cases we have used $ \alpha_1=0 $  and $ \alpha_2 =  - \pi/2 $, in order to obtain 
four well separated initial soliton dips. 
In Fig. \ref{figua}, we show snapshots of the density and phase 
for the case $ X_1= 0.2 $ and $ X_2=0.4$, where the top panels correspond to the initial configuration
and the bottom panels illustrate how the phase remains almost
 constant at both sides of each density dip during the  evolution, aside from small fluctuations associated to
sound excitations.
In particular, it may be seen at the bottom panels that the slower solitons, which exhibit the
deeper density notches, have performed half a cycle around the torus, while the faster solitons have almost
completed an entire one. We note that such a time turns out to be about 14\% smaller than the one corresponding
to the same trajectory of noninteracting  solitons.

It is worthwhile noticing that initial states of this kind can be experimentally achieved by 
using a phase imprinting
method consisting in illuminating the bottom ($y<0$) and left ($x<0$) half-spaces with different intensities,
where, in the case of the top panel of Fig. \ref{figua}, the former intensity has been assumed to be 
higher than the latter. The impression of such a distribution of phases can be  experimentally implemented 
by using a Spatial Light Modulator (SLM) \cite{perrin} with
different intensities in each quadrant of the $ (x,y)$-plane that fulfill the above mentioned condition.

\begin{figure*}
\centering
\includegraphics{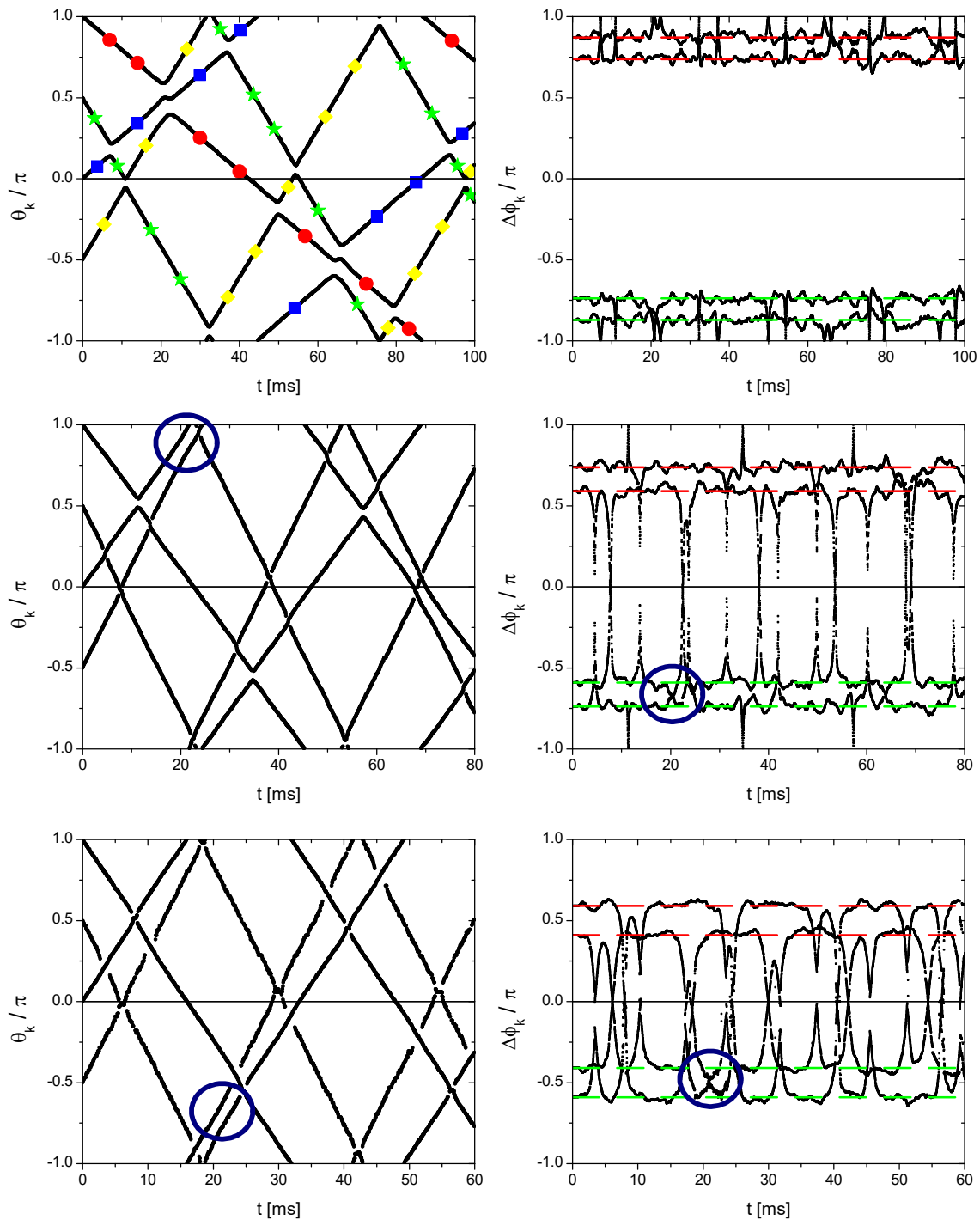}
\caption{Time evolutions of the soliton angular positions (left panels)
and phase differences at both sides of each density minimum (right panels),
for the  initial order parameter (\ref{2gaus}) with  $ \alpha_1=0 $, 
$ \alpha_2 =  - \pi/2 $, and $ X_1=0.2$  and $ X_2 = 0.4$ 
 ( top panel ),
 $ X_1=0.4$ and  $ X_2 = 0.6$ (middle panel), and  $ X_1=0.6$  and $ X_2 = 0.8$ (bottom panel).
The horizontal red (green) dashed lines on the right panels indicate the initial phase difference given 
by Eq. (\ref{phasedif}) for the $j(-j)$-soliton. The empty blue circles in the middle and bottom panels
locate the overtaking collisions mentioned in the text. Circles, squares, stars, and
diamonds are depicted at the top left panel to guide the eye on the different quasiparticle trajectories. 
The symbols have the same color of the arrows that distinguish each soliton
at the left panels of Fig. \ref{figua}.}
\label{figu4}
\end{figure*}

The numerical evolutions for three initial arrays are depicted in Fig. \ref{figu4}, where we show
 the ($k$)-soliton angular position  $\theta_k(t)$ (left panels)
 and  the corresponding phase difference $\Delta \phi_k$ between both  sides of the ($k$)-soliton (right panels).
We are assigning the prefix value $k = j (-j)$ to the soliton with positive (negative) angular velocity 
imprinted through the parameter $X_j$.  
 In such arrays,  the combination of values of the parameters $ X_j$ have been chosen to
cover the distinct behaviors at collisions.
It is important to  observe that  in all cases, each 
soliton moves with a constant angular velocity, except near collisions, as  may be seen in the left
 panels of Fig. \ref{figu4}. 
   Hence, the   quasiparticle picture for describing the dynamics of our soliton  system
 turns out to be plausible. We note that the mass associated with a soliton depends on the
depth of the soliton density notch, which in turn depends on its velocity. Then, we may say that each 
soliton behaves as a quasiparticle with a fixed mass.
 In contrast to the configuration studied in Ref. \cite{je16}, all the present cases have shown that solitons can 
undergo many collisions
without evidencing any signature of energy dissipation that could
 be inferred from a velocity increase.

It is worthwhile mentioning that in an experimental work on
bright solitons \cite{ngu14}, it has been shown that in an asymmetric collision, the solitons pass
through one another and emerge from the collision unaltered
in shape, amplitude, or velocity, but with a new trajectory. It is important to remark  that this behavior
 is also observed for dark solitons 
in all the simulations we have performed, as can be seen in Fig. 5. In contrast, such an effect cannot 
be discerned in the simulations 
of Ref. \cite{je16}, where only symmetric collisions take place and hence each soliton cannot be
distinguished. 

We notice that during an asymmetric collision the precise location of the faster 
soliton could eventually become undetermined, if its density minimum disappears at merging with the 
deeper density minimum of the slower soliton, when the distance between both density dips becomes 
smaller than two healing lengths. This is the case for the faster solitons of the
middle and bottom left panels of Fig. \ref{figu4}.

The imprinted phase difference can be estimated by considering 
the asymptotic values $ \pm 1$ of the hyperbolic tangent in Eq. (\ref{2gaus}), which determine
the phase at each side of the soliton density notch as the phases of the complex numbers 
 $   \pm \sqrt{1- 
X_j^2}   + i X_j  $, yielding the initial  phase difference
\begin{equation}
\Delta \phi_{\pm j} = \mp \,  2 \cos^{-1} (X_j ) 
\label{phasedif}
\end{equation}
for the ($\pm j$)-soliton. It is worthwhile noticing  that  the above expression can be used in an experiment to  
 obtain the parameters $X_j $ by using the values of the  imprinted  local phases.

In the right panels of Fig. \ref{figu4},  we show the initial values obtained from Eq. (\ref{phasedif})
as horizontal dashed lines. Note that the mean value of  each  phase
 difference remains around its initial value,  except near collisions. Such a behavior is also a signature
of the lack of dissipation in the soliton system, as the phase difference is directly linked to the
soliton velocity through Eq. (\ref{phasedif}). 
In particular, from the time evolutions provided by the simulations in Ref. \cite{je16}, it may be 
seen that whenever
 the main absolute value 
of the phase difference decreases,  the absolute value of the velocity increases.

Taking into account the head-on collisions, we can distinguish the following characteristics.
 In the top right panel of Fig. \ref{figu4}, we may observe
  that the phase differences  at 
collisions  reach $\pm \pi$ for  $ X_1=0.2$  and $ X_2 = 0.4$, which
 is in accordance with that observed in Ref. \cite{je16}
 for velocities slower 
than half the sound speed ( $X_j<0.5$).
On the other hand, for $X_1=0.6$ and  $ X_2=0.8$ (bottom panel), we have both soliton velocities above
such a limit and 
the phase differences vanish 
at collisions, remaining bounded along the whole evolution, 
as previously observed for symmetric collisions \cite{je16}. Finally, 
we may observe in the middle right panel that  both behaviors coexist for $X_1=0.4$ and $ X_2= 0.6$,
as expected.

A different class of soliton interaction, arising only for asymmetric collisions, 
takes place when an overtaking event
occurs.
Such overtaking collisions can be viewed for example in the middle and bottom panels of Fig. \ref{figu4},
indicated by the empty circles at $t \simeq 20$ ms  and 
$t \simeq 22$ ms, respectively. Note also in the right panels, the phase difference 
approaches that can be observed in 
between adjacent horizontal dashed lines for this kind of collisions, in contrast to the above high
 velocity head-on collisions, where the phase differences go to zero.

\subsection{Quasiparticle picture}

%\subsubsection{Equations of motion}

In this subsection we will treat the solitons as quasiparticles. In this picture, each quasiparticle  has  a fixed (negative)
mass and
can freely move in the angular direction. The interaction between such quasiparticles occurs 
through a particular type of 
collision, where both colliding quasiparticles are transmitted through each other without altering their velocity. As a consequence
of such a collision, an angular shift on each soliton position is produced. 
We recall that this behavior has 
been experimentally observed for bright solitons \cite{ngu14}. 

By means of the GP numerical simulations, we have shown that the velocity of each soliton remains
 constant, except around collisions. As a consequence, the depth of the soliton density dip 
is conserved, which implies that
the soliton mass is also conserved. Hence, the quasiparticle description can be safely applied
and the trajectory $\theta_k(t)$ of a soliton with angular velocity 
$ \omega_k $ can be written as,
\begin{equation}
\theta_k(t)= \omega_k  t  + \sum_l  \Delta \theta_k(t_l) \, \,  {\cal U}(t-t_l) + \theta_k(0) ,
\label{traje}
\end{equation}
where ${\cal U} (t-t_l) $
denotes the unit step function, $t_l$ indicates the time of the $l$-th collision,
 and   $\Delta \theta_k(t_l)$ designates the angular shift  produced at such a collision.

We note that the number of collisions,  in contrast to the 1D infinite case, is not upper bounded 
and could take any value, depending
on the time interval involved. However, we will see that the shifts could only take a few  
different values.

As seen in the previous section,
we have considered initial configurations with two distinct values of the parameter $X_j$, which gives rise to
a system of four rotating solitons, each one identified by its corresponding angular velocity $ \omega_k$.
We will see that such a system could  yield only  six absolute values of the shift.
In the next subsections, we will derive a relation between the shifts involved in an asymmetric collision,
and we will also evaluate all of them using a numerical method and an analytical approach.

\subsubsection{Theoretical treatment of the shifts}

In an infinite  1D system, the spatial shifts $\Delta z_i$ at a collision
 have been analytically  obtained using the explicit form of the order parameter \cite{zak73} as,
\begin{eqnarray}
 |\Delta z_i| & = & \frac{ \zeta }{ 2  \sqrt{1- Y_i^2}}\nonumber\\
& \times &    \ln
\left[   \frac{  \left(  Y_i - Y_j  \right)^2 + \left( \sqrt{ 1- Y_i^2} + 
\sqrt{ 1- Y_j^2} \right)^2  }{  \left(  Y_i - Y_j \right)^2 + \left( \sqrt{ 1- Y_i^2} - 
\sqrt{ 1- Y_j^2} \right)^2  } \right], \label{shiftz1}
\end{eqnarray}
where  $i=1,2$ and  $j=1,2$,   with $i \ne j$. Here $Y_k = \dot{z}_k/c $ represents
the velocity (in units of the sound speed $c$) of the corresponding soliton
and $\zeta $ denotes the healing length of the homogeneous system. Both shifts have always opposite signs and,
as previously demonstrated   by Tzusuki \cite{tsu71}
 by analyzing the motion of the soliton mass center during a collision,  they fulfill the following relation
\begin{equation}
 \Delta z_1  \sqrt{1- Y_1^2} +  \Delta z_2  \sqrt{1- Y_2^2}   = 0.
\label{relshiftz12}
\end{equation}
In our toroidal configuration, the hypothesis of
conservation of the linear momentum does not remain valid, however, 
 by applying the  conservation of the angular momentum,
we will
see that an analogous expression to (\ref{relshiftz12}) can still hold.
 We will assume
solitons as point-quasiparticles with masses $M_i$  that move at a fixed radius $r_0$. 

Considering a  time interval  ($0 \le t \le  t_f$)   where a single 
collision takes place, which occurs between the  $(1)$- and $(2)$-soliton  at  $ t= t_c < t_f $,  and  using
Eq. (\ref{traje}), we obtain both  soliton  angular velocities,
\begin{equation}
\dot{\theta}_1(t)= \omega_1   +    \Delta \theta_{1,2} \, \,  \frac{d}{dt}  {\cal U}(t-t_c)  ,
\label{traje1}
\end{equation}
\begin{equation}
\dot{\theta_2}(t)= \omega_2   +  \Delta \theta_{2,1} \, \,  \frac{d}{dt}  {\cal U}(t-t_c)  ,
\label{traje2}
\end{equation}
where  $ \Delta \theta_{1,2} =\Delta \theta_{1}(t_c) $ denotes the angular shift on 
$  \theta_{1}$ produced by the collision with the ($2$)-soliton. Hereafter, all the angular shifts
 will be identified by the subscripts of the solitons involved
in the corresponding collision.

Multiplying the above angular velocities  by the square of the  radius $r^2_0$ and the corresponding 
soliton mass $M_i$,
 we can write the  expressions for 
each soliton angular momentum as,
\begin{equation}
\vec{L}_1=  M_1 r^2_0  \, \omega_1  \hat{z} 
  + \frac{ d}{dt} \left[   M_1 r^2_0  \Delta \theta_{1,2} \, \,  {\cal U}(t-t_c)  \right] \hat{z} ,
\label{l1}
\end{equation}
\begin{equation}
\vec{L}_2=  M_2 r^2_0  \, \omega_2  \hat{z}  
 + \frac{ d}{dt}\left[   M_2 r^2_0  \Delta \theta_{2,1} \, \,  {\cal U}(t-t_c)  \right] \hat{z} ,
\label{l2}
\end{equation} 
where $\hat{z}$ denotes the $z$-coordinate unit vector.
Now, assuming that the total angular momentum $ \vec{L}=\vec{L}_1 + \vec{L}_2$ must be conserved,
the sum of the second terms  of   Eqs. (\ref{l1}) and (\ref{l2}), should at least  be bounded.
Rearranging such a sum we have,
\begin{equation}
\frac{ d}{dt}\left[   r^2_0  ( M_1  \Delta \theta_{1,2} \, 
+  M_2   \Delta \theta_{2,1}) \, \,  {\cal U}(t-t_c)  \right] 
\label{c1}
\end{equation}
and hence, applying to this quantity the condition of being bounded leads to,
\begin{equation}
    M_1   \Delta \theta_{1,2}  +    M_2  \Delta \theta_{2,1}  = 0.
\label{c2}
\end{equation}

On the other hand, the negative mass of the soliton \cite{becker08,mass1} can be estimated by using 
 Eq. (\ref{2gaus})
to calculate the number of
particles expelled from the core, yielding
\begin{equation}
    M_i     =  - 2  m A^2 \frac{\sqrt{1-X_i^2}}{k} \int  dr \frac{r}{r_0}\, \exp \left[ - \gamma \left(1-\frac{r}{r_0}\right)^2\right] ,
\label{mass}
\end{equation}
which replaced in (\ref{c2}) leads to the following expression analogous to (\ref{relshiftz12}),
\begin{equation}
 \Delta \theta_{1,2}  \sqrt{1- X_1^2} +  \Delta \theta_{2,1}  \sqrt{1- X_2^2}   = 0
\label{relshiftteta12}
\end{equation}
for our toroidal configuration.

Given that our condensate is not a 1D ring, we do not  have analytical solutions of our solitonic system, 
and hence we cannot derive an analogous
formula to (\ref{shiftz1})
for the shift values. Nevertheless, in the next section we will show that accurate values can be
obtained by adapting (\ref{shiftz1})
 in a convenient manner.
We want to note that, being our system non homogeneous, the way we adopt  for evaluating the
healing length becomes crucial to guarantee such an accuracy in the values of the shifts.

\subsubsection{Numerical and analytical determination of the angular shifts \label{num}}
\begin{figure*}
\centering
\includegraphics{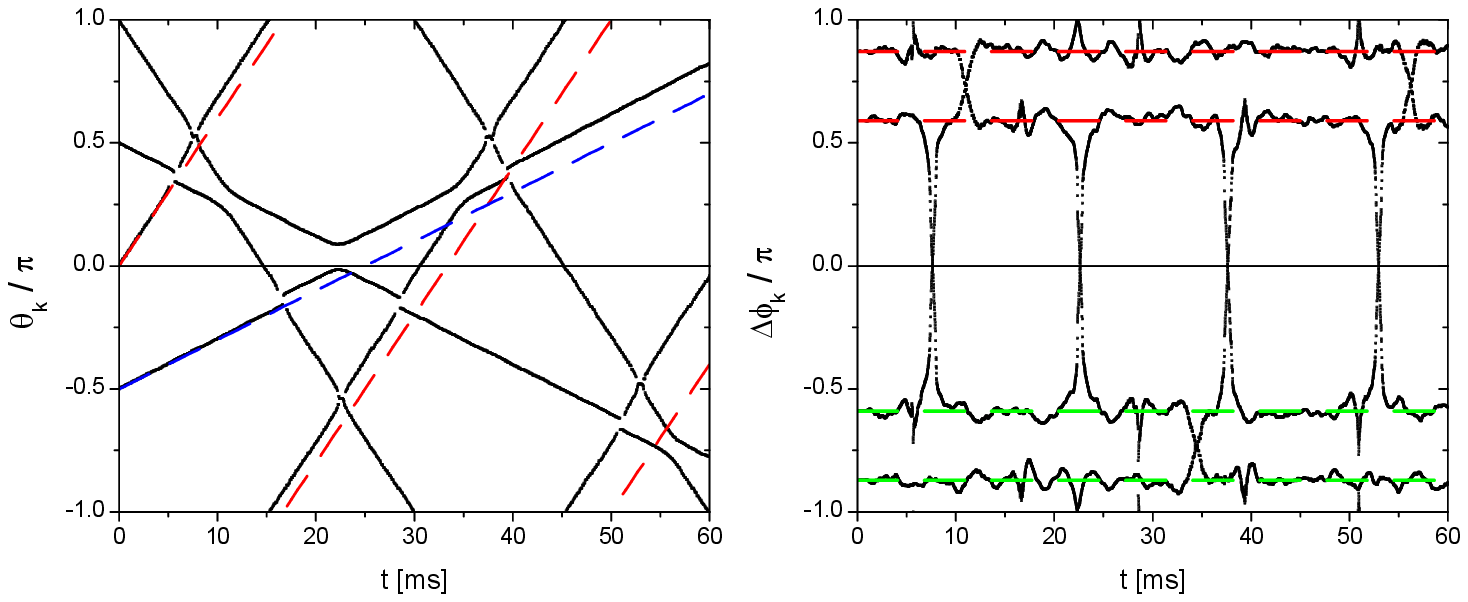}
\caption{Same as Fig. \ref{figu4} for  $ X_1=0.6$  and  $ X_2 = 0.2$. 
The dashed lines in the left panel indicate noninteracting trajectories for the (1)-soliton (red)  
and the (2)-soliton (blue).}
\label{figu8}
\end{figure*}

In order to numerically determine the angular shifts, we have run a GP simulation
for an initial condition with $X_1=0.6$ and $X_2=0.2$, where, as seen in
Fig.~\ref{figu8}, it is clearly shown that there exist four different types of collisions.
By measuring each slope of the four numerically obtained trajectories $\theta_k(t)$ ($k=\pm 1, \pm 2$), 
we have determined the soliton angular velocities
$ \omega_{1} = (0.060 \pm 0.001)\, \pi  \, \rm{ms}^{-1} $ and
$ \omega_{2} = (0.020 \pm  0.001)\, \pi  \, \rm{ms}^{-1} $, being $\omega_{-k}=-\omega_k$.
Here it is important to notice that,
in  analogy  to the infinite 1D case, where   $X_j= v_j/ c$, with $v_j$ the soliton (linear) speed,
we may write $ c(X_1,X_2)= \omega_{1} r_0 /X_1=  \omega_{2} r_0 /X_2 $.
Thus, we can utilize such a proportionality to extract an 
estimate of the linear speed of sound azimuthally propagating along our ring-shaped condensate, yielding
$ c \simeq \,1.26 \,\mu$m/ms. 
%  a value which  lies in between the theoretical estimates for
%a quasi-1D system   aca
% $ \sqrt{g n/ 2 m} = 1.070\, \mu$m/ms and that of a homogeneous system
%$ \sqrt{g n/ m} = 1.513\, \mu$m/ms, where $n$ denotes the density maximum of our initial configuration.
\begin{figure*}
\centering
\includegraphics{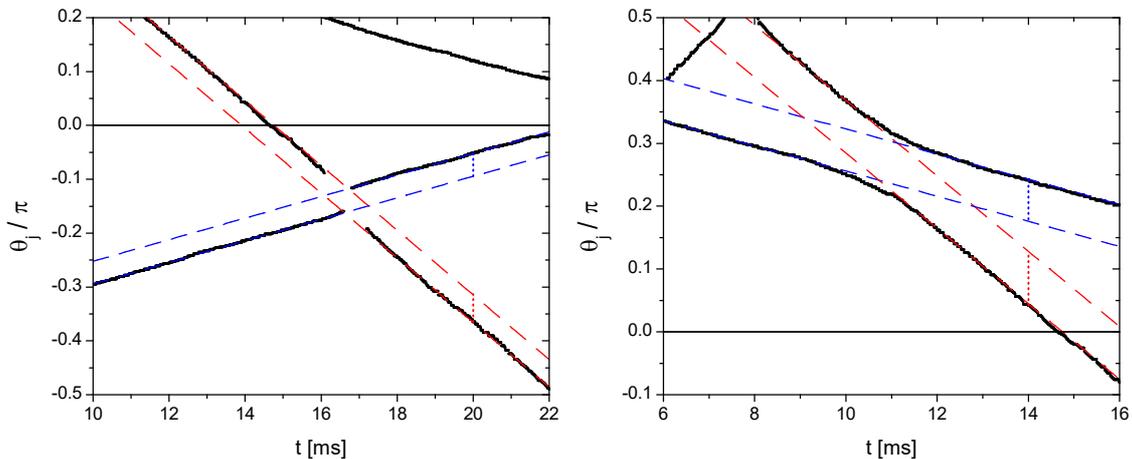}
\caption{Enlarged graphs of both kinds of asymmetric collisions represented at the left panel
of Fig. \ref{figu8}. The left (right) panel shows a head-on (overtaking)  collision with
dashed lines tangent to the trajectories before and after the collisions, while the vertical dotted lines  
indicate the corresponding shifts.}
\label{figu9}
\end{figure*}

To illustrate the way we have calculated the angular shifts, in
Fig. \ref{figu9} we show close-ups of  both kinds of asymmetric
collisions displayed at the left panel of the previous figure, 
drawing the tangent lines to the soliton trajectories before and after the collisions,
with slopes corresponding to the aforementioned angular velocities. This in turn determines the shifts 
denoted by
the vertical dashed lines.
Due to the Cartesian grid, the error in the determination of  the angular shifts
is about $ 0.003 \pi$.
Particularly, for the head-on collision \cite{stellmer08,collision} which involves the
($-1$)- and ($2$)-soliton, depicted in 
the left panel, we obtained for the angular shift on the $(-1)$-soliton $ |\Delta \theta_{-1,2}|=  0.051  \pi $ 
and for the angular shift on the $(2)$-soliton
$ |\Delta \theta_{2,-1}|= 0.042   \pi $.
 So, the quotient turns out to be
$ |\Delta \theta_{2,-1} / \Delta \theta_{-1,2}|=  0.82  \pm 0.07 $, which yields a good agreement with 
the result \\
$\sqrt{1- X_1^2}/ \sqrt{1- X_2^2}= 0.81$, in accordance with Eq. (\ref{relshiftteta12}).
We note that a collision between the  $(1)$- and the $(-2)$-soliton yields the same preceding absolute values
 of the angular shifts,
i.e,  $ |\Delta \theta_{1,-2}|=  |\Delta \theta_{-1,2}|$ 
  and   $ |\Delta \theta_{-2,1}|=|\Delta \theta_{2,-1}|$.

 On the other hand, for the overtaking collision shown on the right panel of Fig. \ref{figu9} that
involves the ($-1$)- and ($-2$)-soliton,
we obtained $ |\Delta \theta_{-1,-2}|= 0.084  \pi $ and \\
$ |\Delta \theta_{-2,-1}|= 0.067   \pi $,
  which yields a quotient \\
$| \Delta \theta_{-2,-1} / \Delta \theta_{-1,-2}|=  0.80 \pm 0.07 $ that again equals \\
$\sqrt{1- X_1^2}/ \sqrt{1- X_2^2} $,
 within the predicted error.  Once more, the ($1$)- and ($2$)-soliton would yield the same shift values.
   
We have calculated in a similar fashion the shifts in the symmetric collisions,
which yielded $ |\Delta \theta_{1,-1}|= |\Delta \theta_{-1,1}| =0.030   \pi $ and $ |\Delta \theta_{2,-2}|= |\Delta \theta_{-2,2}| = 0.076   \pi $.

In summary, for our particular  four-soliton system  only  
six different absolute values of the angular shifts are obtained, which are given in Table 1. In particular, 
we have  two symmetric collisions,
 each one involving a single absolute value of the shift,
 and two types of asymmetric collisions which give rise to the remaining four absolute values. 

In view of the analogy between Eqs. (\ref{relshiftz12}) and (\ref{relshiftteta12}), it becomes evident that
one could also try to adapt formula (\ref{shiftz1}) to our present toroidal  geometry. Thus, taking into account
 that the parameters $X_j$ are
defined as positive quantities, 
we may obtain the absolute values of the different angular shifts as,
\begin{eqnarray}
& &|\Delta \theta_{-i,\pm j}|/ \pi = \frac{   \zeta^{*} / (\pi r_0) }{  2 \sqrt{1- X_i^2}}\nonumber\\
&&\times   \ln
\left[   \frac{  \left(  X_i \pm X_j  \right)^2 + \left( \sqrt{ 1- X_i^2} + 
\sqrt{ 1- X_j^2} \right)^2  }{  \left(  X_i \pm X_j \right)^2 + \left( \sqrt{ 1- X_i^2} - \sqrt{ 1- X_j^2} \right)^2  } \right],
\label{shift1}
\end{eqnarray}
where  $i=1,2$ and  $j=1,2$. The sign $ + $ ($-$) corresponds to head-on (overtaking) collisions.
Here it is important to recall that, as our system is non homogeneous, the value of the healing
length would depend on the density of the selected point at which it is calculated. Conversely,  the value of the 
sound speed, obtained in the previous section, only depends on well defined quantities:  $X_j$ and the ($j$)-soliton velocity.
Hence,
we have utilized the expression $ \zeta^{*} =  \hbar/(m c )$ for calculating the healing length with
our above estimate of the sound speed, $ c = \,1.26 \,\mu$m/ms, which yields $ \zeta^{*} =0.575\, \mu$m.
 The corresponding results arising from Eq. (\ref{shift1}),
 which agree very well with those obtained from numerical simulations, 
are summarized in Table \ref{tab1}. Also it is worth noting that the use of
other common estimates of the healing length, such as  that derived for 
a homogeneous
system at the density maximum, would have yielded an
underestimated shift in about a 20 percent.
\begin{table}
\caption{ Absolute value of the angular shifts (in units of $\pi$) produced at the collisions. }
\begin{tabular}{lcccc}
\hline\noalign{\smallskip}
Type of collision   &   $ X_k $  &  Simulation    &     Eq. (\ref{shift1})   &  $  $  \\
\noalign{\smallskip}\hline\noalign{\smallskip}
 Head-on asymmetric  &  $  0.6  $ &   $  0.051   $ &  $ 0.050    $ &  $      $  \\
 &  $  0.2 $ &    $    0.042  $ &   $  0.041    $ &  $      $ \\
\hline
 Overtaking asymmetric &  $  0.6  $ &   $  0.084  $ &  $ 0.082     $ &  $      $  \\[3pt]
  &  $  0.2 $ &    $    0.067 $ &   $  0.067    $ &  $      $ \\[3pt]
\hline
Head-on symmetric   &  $  0.6  $ &   $  0.030  $ &  $ 0.029    $ &  $      $  \\[3pt]
   &  $  0.2 $ &    $    0.076 $ &   $  0.076   $ &  $      $ \\[3pt]
\noalign{\smallskip}\hline
\end{tabular}
\label{tab1}
\end{table}

\subsubsection{Sound velocity}\label{sound}
\begin{figure*}
\centering
\includegraphics{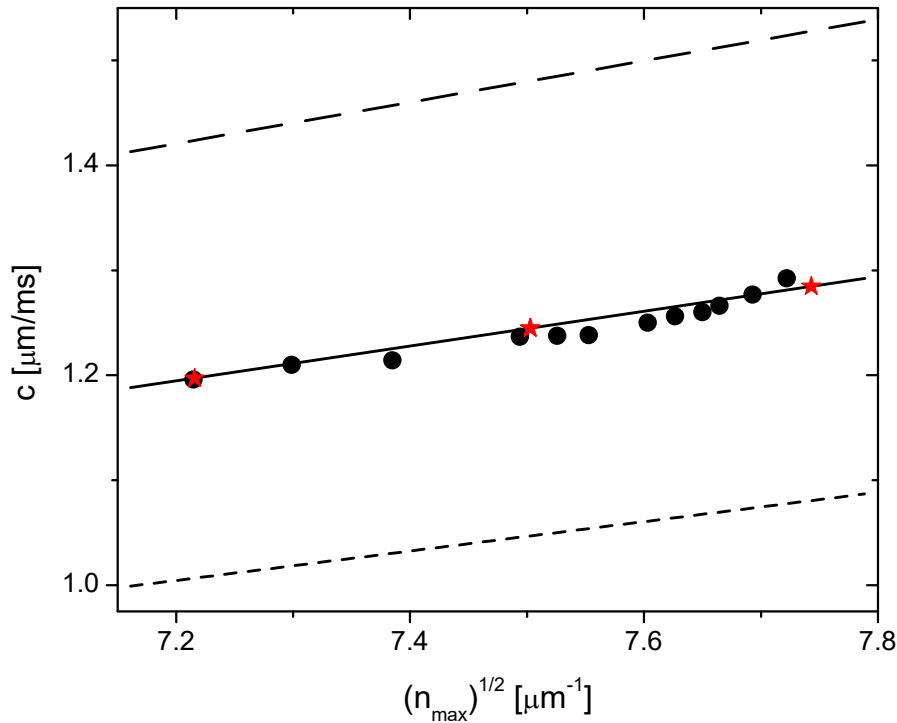}
\caption{Sound velocity as a function of the square root of the density maximum $n_{max}$. 
The solid  line corresponds to $ c=\sqrt{g n_{max} / \sqrt{2}m} $, which has been extracted by using the
Gaussian density profile.
The dashed (dotted) line corresponds to $ c=\sqrt{g n_{max} / m} $ ($ c=\sqrt{g n_{max} / 2 m} $).
The black solid circles  have been obtained from GP simulations in the manner   outlined in Sec. \ref{num}. 
The red stars correspond to the sound velocity derived in Sec. \ref{sound}, with 
 the densities  obtained from Eq. (\ref{nmax}) for
 $(X_1,X_2)= (0.9, 1), (0.7,0.8)$, and $ (0.1,0.2) $, from left to right.
}
\label{figuc}
\end{figure*}

Given the importance that the sound velocity acquires for determining the angular shifts,
we have obtained its numerical values when varying the imprinted phases by using the
relationship $c(X_1,X_2)= \omega_j  r_0/ X_j$. Such values are depicted as solid circles
 in Fig. \ref{figuc}, as a function of the square root of 
the soliton density maximum.  
On the other hand,  we have theoretically derived an analytic formula for the sound velocity in terms of 
our soliton 2D density maximum.
With such a purpose, we have used  the result that the sound propagation speed in an elongated condensate reads 
$ c=\sqrt{g \bar{n}/ m} $ \cite{kavoulakis98},
 where $\bar{n}$ is the averaged density in the  transversal direction.
 Then, we have considered  a 2D elongated condensate with a density profile in the transversal direction
defined by the coordinate $r$ that
emulates the  density profile of our solitonic system far from the density notches, which, using (6)
can be modeled as,
\begin{equation}
n(r)=  n_{max} \, \exp \left[ - \gamma\left(1-\frac{r}{r_0}\right)^2\right],
\label{grounds}
\end{equation}
where $n_{max} $ denotes the density maximum. Such a maximum changes with the imprinted phases
and hence it depends on $X_1$ and $X_2$. An analytical expression for \\
$n_{max}(X_1,X_2)$ will be derived
below.
Using the above profile, we have calculated the mean value $\bar{n}=\int n^2 dr / \int n dr $,  which yields 
 $\bar{n}= n_{max}/\sqrt{2}$ and hence $ c=\sqrt{g n_{max}/ \sqrt{2} m} $.
It is worthwhile  recalling  that in the three-dimensional case, the value $\bar{n}= n_{max}/ 2$ 
is obtained \cite{kavoulakis98}.
It can be seen in Fig. 8  that the curve  $ c=\sqrt{g n_{max}/ \sqrt{2} m} $ is in a  
very  good accordance with the set of points obtained from the time dependent simulations.
On the other hand, the upper curve  $ c=\sqrt{g n_{max}/  m} $     represents  
the velocity that should be used to calculate the healing length by employing  its
analytical expression as a function of the local density $   \zeta  = \hbar / \sqrt{n_{max} g m} $. 
For completeness we have also  depicted the function
 $ c=\sqrt{g n_{max}/ 2 m} $ (lower line),  which represents the expression for a three-dimensional density. 

As we are dealing with a fixed total  number of particles, the value of  $n_{max}$ increases with
the imprinted phase difference. In particular, $n_{max}(X_1,X_2)$ aquires its maximum value when $ X_1=X_2=0$,
which corresponds
to a black  soliton. We can go further and obtain an analytical expression for such a quantity
by evaluating the amount of particles expelled
from the place where each soliton density notch is formed, instead of using $n_{max}=A^2$ of Eq. (6). 
Thus we obtain,
\begin{equation}
 n_{max}(X_1,X_2) = \frac{n_0}{1-\frac{2\sqrt{1-X_1^2}}{ \pi   {\rm k} r_0} -\frac{2\sqrt{1-X_2^2}}{ \pi   {\rm k}  r_0}}
\label{nmax}
\end{equation}
where  $n_ 0  = n_{max}(X_1=1,X_2 =1)  $ is the ground state density maximum. 
We have found that with
such an  estimate, the sound velocity is obtained within a relative error of  $ 2 $ percent. 
We have plotted as red stars in Fig. 8 some representative points obtained by using this simplified 
protocol. The nearest black solid circles, obtained from the time dependent simulations, correspond
to the same set of parameters $X_1$ and $X_2$. 
From these results, we may conclude that a good quasiparticle description can be achieved 
by solely using the ground state density
 and the imprinted phases. It is worthwhile noticing that,
given that the experimental images correspond to integrated densities along the direction through which the
condensate is viewed, yielding thus images proportional to  2D densities,   
our result could also be applicable to such  configurations. 

\subsubsection{Accumulative shift effect}
 Here it is important to note that, although 
each individual shift at a collision may not be expected to lie within the reach of the 
available experimental resolution 
\cite{stellmer08}, an
 accumulative effect of the shifts during one period could in principle be experimentally detectable.
The magnitude of such an effect can be viewed in Fig. \ref{figu8}, where 
one can observe that the collisions cause  a sizable advance on the soliton 
dynamics, 
as compared to the trajectories of ideally noninteracting solitons, which are depicted as dashed lines.
We can analytically estimate this effect by using the velocities and the angular shifts. For instance,
 the period for the noninteracting trajectory starting at $\theta=0$ turns out to be 33.3 ms,
 whereas the shifts produced at each collision bring about that this time becomes effectively reduced.
In particular, we may see that  the corresponding  (1)-soliton undergoes two asymmetric and two
 symmetric collisions 
during such a time interval,
with an estimated reduction in time $ |\Delta \theta_{1,k} /\omega_1|$ at each collision with another 
 $(k)$-soliton, 
which yields a reduction in the whole period of about 2.7 ms ($\sim 10\%$).
On the other hand, for the trajectory that starts at $\theta= - \pi /2$  that corresponds to the (2)-soliton, 
we obtain that its 
 half-period becomes reduced in 4.7 ms with respect to the noninteracting one (50 ms),
which means again about a $10\%$.
It is worthwhile noticing  that only in this case a negative shift is produced in the evolution, which comes
from  the
overtaking collision ( see left panel 
of Fig. \ref{figu8} for $t\simeq 35 $ ms).
Such analytical calculations are in accordance with the deviations between the trajectories of 
 noninteracting solitons and GP simulations observed in Fig. \ref{figu8}. 

\section{Conclusions}\label{sec5}

By choosing a suitable trapping potential, 
we have furnished a toroidal condensate, whose soliton dynamics is 
appropiate for being described as a
 quasiparticle system with constant energy.
More specifically, each soliton trajectory can be described  
 in terms of a constant velocity plus the angular shifts produced at each collision.

The soliton states are created by proposing a variational ground state Gaussian profile in  the
radial direction and next imprinting
phase differences through the parameters $X_j$, with the same structure in the angular direction
 as that 
 obtained for infinite 1D  systems.
In fact, we have shown that introducing different configurations with such initial  states 
into the time dependent GP equation, the solitons
evolve conserving their  velocities except around collisions. 

In order to analyze the values of the angular shifts involved in the dynamics,
we have numerically determined by means of GP simulations 
the six different shifts arising in a four-soliton system.
In addition, 
we have derived in a first step a relationship between the shifts produced in a particular  collision,
assuming both solitons as massive quasiparticles that collide conserving the angular momentum.
Next, we show that very accurate shift values can be analytically obtained by adapting the corresponding 
expression of an infinite 1D system
to our present toroidal geometry. It is important to recall that such a formula depends on the healing length,
whose determination in an inhomogeneous system becomes vague.
However, we have successfully overcome this drawback by using  the sound
velocity propagating in the angular direction.
We have shown that the sound velocity  
can be calculated either by employing the relationship between the soliton velocity and the 
parameter $X_j$, or by using a prescription we have derived, 
which employs solely the ground state density.
Such a procedure have yielded very accurate results, 
as compared to those arising from GP simulations.

We emphasize the fact that the value of the  sound speed
azimuthally propagating along the toroidal condensate acquires a crucial role in our calculations, so we have
paid special attention to its determination.

To conclude we first recall that  toroidal  condensates 
are used in current experiments \cite{seaman,boshier13,jen14,eckel}.
We  believe that the phase imprinting method that we have outlined in Sec. \ref{GPsim},
 which consists in 
simultaneously illuminating different half-spaces with distinct intensities,
could be experimentally applied. 
Then, using Eq. (\ref{phasedif}) the parameters $X_j= \cos(\Delta \phi_{j}/2)$ 
could be extracted from the imprinted phases.
On the other hand, by fitting the  2D ground state density obtained from experimental images
with a Gaussian profile and following the procedure explained
in Sec. \ref{sound},
 the sound velocity could be accurately determined for each set of imprinted phases, as shown
in Fig. \ref{figuc}.
Therefore, having the values of the sound velocity $c$ and the parameters $X_j$, the quasiparticle picture 
should be completely defined.

Finally, in case that the individual shifts could not be measured for lying out of 
the current experimental resolution 
\cite{stellmer08}, we have proposed to determine 
them by using  long enough evolutions, where the shift values are accumulated.
To conclude such a process,  just a set of linear
equations should remain to be solved.
  In particular, we have shown that in our simulations the soliton trajectories turn out to be advanced
in about 
a $10\%$ of a time period  
with respect to the ideally noninteracting ones.

\begin{acknowledgement}
  We acknowledge CONICET for financial support under Grant
PIP 11220150100442CO. HMC acknowledges Universidad de Buenos Aires for financial support 
under Grant UBA-CyT 20020150100157.
\end{acknowledgement}

\subsection*{Author contribution statement}
Both authors were equally involved in the preparation of the manuscript.

%\bibliography{paper}

\begin{thebibliography}{34}
\expandafter\ifx\csname natexlab\endcsname\relax\def\natexlab#1{#1}\fi
\expandafter\ifx\csname bibnamefont\endcsname\relax
  \def\bibnamefont#1{#1}\fi
\expandafter\ifx\csname bibfnamefont\endcsname\relax
  \def\bibfnamefont#1{#1}\fi
\expandafter\ifx\csname citenamefont\endcsname\relax
  \def\citenamefont#1{#1}\fi
\expandafter\ifx\csname url\endcsname\relax
  \def\url#1{\texttt{#1}}\fi
\expandafter\ifx\csname urlprefix\endcsname\relax\def\urlprefix{URL }\fi
\providecommand{\bibinfo}[2]{#2}
\providecommand{\eprint}[2][]{\url{#2}}

\bibitem[{\citenamefont{Denardo et~al.}(1990)\citenamefont{Denardo, Wright,
  Putterman, and Larraza}}]{denardo}
\bibinfo{author}{\bibfnamefont{B.}~\bibnamefont{Denardo}},
  \bibinfo{author}{\bibfnamefont{W.}~\bibnamefont{Wright}},
  \bibinfo{author}{\bibfnamefont{S.}~\bibnamefont{Putterman}},
  \bibnamefont{and} \bibinfo{author}{\bibfnamefont{A.}~\bibnamefont{Larraza}},
  \bibinfo{journal}{Phys. Rev. Lett.} \textbf{\bibinfo{volume}{64}},
  \bibinfo{pages}{1518} (\bibinfo{year}{1990}).

\bibitem[{\citenamefont{Chen et~al.}(1993)\citenamefont{Chen, Tsankov, Nash,
  and Patton}}]{chen}
\bibinfo{author}{\bibfnamefont{M.}~\bibnamefont{Chen}},
  \bibinfo{author}{\bibfnamefont{M.~A.} \bibnamefont{Tsankov}},
  \bibinfo{author}{\bibfnamefont{J.~M.} \bibnamefont{Nash}}, \bibnamefont{and}
  \bibinfo{author}{\bibfnamefont{C.~E.} \bibnamefont{Patton}},
  \bibinfo{journal}{Phys. Rev. Lett.} \textbf{\bibinfo{volume}{70}},
  \bibinfo{pages}{1707} (\bibinfo{year}{1993}).

\bibitem[{\citenamefont{Heidemann et~al.}(2009)\citenamefont{Heidemann,
  Zhdanov, S{\"{u}}tterlin, Thomas, and Morfill}}]{heide}
\bibinfo{author}{\bibfnamefont{R.}~\bibnamefont{Heidemann}},
  \bibinfo{author}{\bibfnamefont{S.}~\bibnamefont{Zhdanov}},
  \bibinfo{author}{\bibfnamefont{R.}~\bibnamefont{S{\"{u}}tterlin}},
  \bibinfo{author}{\bibfnamefont{H.~M.} \bibnamefont{Thomas}},
  \bibnamefont{and} \bibinfo{author}{\bibfnamefont{G.~E.}
  \bibnamefont{Morfill}}, \bibinfo{journal}{Phys. Rev. Lett.}
  \textbf{\bibinfo{volume}{102}}, \bibinfo{pages}{135002}
  (\bibinfo{year}{2009}).

\bibitem[{\citenamefont{Kivshar and Luther-Davies}(1998)}]{kivshar}
\bibinfo{author}{\bibfnamefont{Y.~S.} \bibnamefont{Kivshar}} \bibnamefont{and}
  \bibinfo{author}{\bibfnamefont{B.}~\bibnamefont{Luther-Davies}},
  \bibinfo{journal}{Phys. Rep.} \textbf{\bibinfo{volume}{298}},
  \bibinfo{pages}{81} (\bibinfo{year}{1998}).

\bibitem[{\citenamefont{Tsuzuki}(1971)}]{tsu71}
\bibinfo{author}{\bibfnamefont{T.}~\bibnamefont{Tsuzuki}}, \bibinfo{journal}{J.
  Low Temp. Phys.} \textbf{\bibinfo{volume}{4}}, \bibinfo{pages}{441}
  (\bibinfo{year}{1971}).

\bibitem[{\citenamefont{Zakharov and Shabat}(1973)}]{zak73}
\bibinfo{author}{\bibfnamefont{V.~E.} \bibnamefont{Zakharov}} \bibnamefont{and}
  \bibinfo{author}{\bibfnamefont{A.~B.} \bibnamefont{Shabat}},
  \bibinfo{journal}{Zh. Eksp. Teor. Fiz.} \textbf{\bibinfo{volume}{64}},
  \bibinfo{pages}{1627} (\bibinfo{year}{1973}).

\bibitem[{\citenamefont{Pitaevskii and Stringari}(2003)}]{libro1}
\bibinfo{author}{\bibfnamefont{L.~P.} \bibnamefont{Pitaevskii}}
  \bibnamefont{and}
  \bibinfo{author}{\bibfnamefont{S.}~\bibnamefont{Stringari}},
  \emph{\bibinfo{title}{Bose-Einstein Condensation}}
  (\bibinfo{publisher}{Oxford University Press}, \bibinfo{address}{Oxford},
  \bibinfo{year}{2003}).

\bibitem[{\citenamefont{Konotop}(2008)}]{libro}
\bibinfo{author}{\bibfnamefont{V.~V.} \bibnamefont{Konotop}},
  \emph{\bibinfo{title}{in Emergent Nonlinear Phenomena in Bose-Einstein
  Condensates}} (\bibinfo{publisher}{Springer-Verlag},
  \bibinfo{address}{Heidelberg}, \bibinfo{year}{2008}).

\bibitem[{\citenamefont{Theocharis et~al.}(2010)\citenamefont{Theocharis,
  Weller, Ronzheimer, Gross, Oberthaler, Kevrekidis, and
  Frantzeskakis}}]{theo10}
\bibinfo{author}{\bibfnamefont{G.}~\bibnamefont{Theocharis}},
  \bibinfo{author}{\bibfnamefont{A.}~\bibnamefont{Weller}},
  \bibinfo{author}{\bibfnamefont{J.~P.} \bibnamefont{Ronzheimer}},
  \bibinfo{author}{\bibfnamefont{C.}~\bibnamefont{Gross}},
  \bibinfo{author}{\bibfnamefont{M.~K.} \bibnamefont{Oberthaler}},
  \bibinfo{author}{\bibfnamefont{P.~G.} \bibnamefont{Kevrekidis}},
  \bibnamefont{and} \bibinfo{author}{\bibfnamefont{D.~J.}
  \bibnamefont{Frantzeskakis}}, \bibinfo{journal}{Phys. Rev. A}
  \textbf{\bibinfo{volume}{81}}, \bibinfo{pages}{063604}
  (\bibinfo{year}{2010}).

\bibitem[{\citenamefont{Frantzeskakis}(2010)}]{rev10}
\bibinfo{author}{\bibfnamefont{D.~J.} \bibnamefont{Frantzeskakis}},
  \bibinfo{journal}{J. Phys. A: Math. Theor.} \textbf{\bibinfo{volume}{43}},
  \bibinfo{pages}{213001} (\bibinfo{year}{2010}).

\bibitem[{\citenamefont{Donadello et~al.}(2014)\citenamefont{Donadello,
  Serafini, Tylutki, Pitaevskii, Dalfovo, Lamporesi, and
  Ferrari}}]{donadello14}
\bibinfo{author}{\bibfnamefont{S.}~\bibnamefont{Donadello}},
  \bibinfo{author}{\bibfnamefont{S.}~\bibnamefont{Serafini}},
  \bibinfo{author}{\bibfnamefont{M.}~\bibnamefont{Tylutki}},
  \bibinfo{author}{\bibfnamefont{L.~P.} \bibnamefont{Pitaevskii}},
  \bibinfo{author}{\bibfnamefont{F.}~\bibnamefont{Dalfovo}},
  \bibinfo{author}{\bibfnamefont{G.}~\bibnamefont{Lamporesi}},
  \bibnamefont{and} \bibinfo{author}{\bibfnamefont{G.}~\bibnamefont{Ferrari}},
  \bibinfo{journal}{Phys. Rev. Lett.} \textbf{\bibinfo{volume}{113}},
  \bibinfo{pages}{065302} (\bibinfo{year}{2014}).

\bibitem[{\citenamefont{Ku et~al.}(2014)\citenamefont{Ku, Ji, Mukherjee,
  Guardado-Sanchez, Cheuk, Yefsah, and Zwierlein}}]{ku14}
\bibinfo{author}{\bibfnamefont{M.~J.~H.} \bibnamefont{Ku}},
  \bibinfo{author}{\bibfnamefont{W.}~\bibnamefont{Ji}},
  \bibinfo{author}{\bibfnamefont{B.}~\bibnamefont{Mukherjee}},
  \bibinfo{author}{\bibfnamefont{E.}~\bibnamefont{Guardado-Sanchez}},
  \bibinfo{author}{\bibfnamefont{L.~W.} \bibnamefont{Cheuk}},
  \bibinfo{author}{\bibfnamefont{T.}~\bibnamefont{Yefsah}}, \bibnamefont{and}
  \bibinfo{author}{\bibfnamefont{M.~W.} \bibnamefont{Zwierlein}},
  \bibinfo{journal}{Phys. Rev. Lett.} \textbf{\bibinfo{volume}{113}},
  \bibinfo{pages}{065301} (\bibinfo{year}{2014}).

\bibitem[{\citenamefont{Chevy}(2014)}]{chevy14}
\bibinfo{author}{\bibfnamefont{F.}~\bibnamefont{Chevy}},
  \bibinfo{journal}{Physics} \textbf{\bibinfo{volume}{7}}, \bibinfo{pages}{82}
  (\bibinfo{year}{2014}).

\bibitem[{\citenamefont{Lamporesi et~al.}(2013)\citenamefont{Lamporesi,
  Donadello, Serafini, Dalfovo, and Ferrari}}]{lamporesi13}
\bibinfo{author}{\bibfnamefont{G.}~\bibnamefont{Lamporesi}},
  \bibinfo{author}{\bibfnamefont{S.}~\bibnamefont{Donadello}},
  \bibinfo{author}{\bibfnamefont{S.}~\bibnamefont{Serafini}},
  \bibinfo{author}{\bibfnamefont{F.}~\bibnamefont{Dalfovo}}, \bibnamefont{and}
  \bibinfo{author}{\bibfnamefont{G.}~\bibnamefont{Ferrari}},
  \bibinfo{journal}{Nature Phys.} \textbf{\bibinfo{volume}{9}},
  \bibinfo{pages}{656} (\bibinfo{year}{2013}).

\bibitem[{\citenamefont{Fedichev et~al.}(1999)\citenamefont{Fedichev, Muryshev,
  and Shlyapnikov}}]{fedi99}
\bibinfo{author}{\bibfnamefont{P.~O.} \bibnamefont{Fedichev}},
  \bibinfo{author}{\bibfnamefont{A.~E.} \bibnamefont{Muryshev}},
  \bibnamefont{and} \bibinfo{author}{\bibfnamefont{G.~V.}
  \bibnamefont{Shlyapnikov}}, \bibinfo{journal}{Phys. Rev. A}
  \textbf{\bibinfo{volume}{60}}, \bibinfo{pages}{3220} (\bibinfo{year}{1999}).

\bibitem[{\citenamefont{Busch and Anglin}(2000)}]{bus00}
\bibinfo{author}{\bibfnamefont{T.}~\bibnamefont{Busch}} \bibnamefont{and}
  \bibinfo{author}{\bibfnamefont{J.~R.} \bibnamefont{Anglin}},
  \bibinfo{journal}{Phys. Rev. Lett} \textbf{\bibinfo{volume}{84}},
  \bibinfo{pages}{2298} (\bibinfo{year}{2000}).

\bibitem[{\citenamefont{Konotop and Pitaevskii}(2004)}]{konotop04}
\bibinfo{author}{\bibfnamefont{V.~V.} \bibnamefont{Konotop}} \bibnamefont{and}
  \bibinfo{author}{\bibfnamefont{L.}~\bibnamefont{Pitaevskii}},
  \bibinfo{journal}{Phys. Rev. Lett.} \textbf{\bibinfo{volume}{93}},
  \bibinfo{pages}{240403} (\bibinfo{year}{2004}).

\bibitem[{\citenamefont{Becker et~al.}(2008)\citenamefont{Becker, Stellmer,
  Soltan-Panahi, {\relax S. D\"orscher}, Baumert, Richter, {\relax J.
  Kronj\"ager}, Bongs, and Sengstock}}]{becker08}
\bibinfo{author}{\bibfnamefont{C.}~\bibnamefont{Becker}},
  \bibinfo{author}{\bibfnamefont{S.}~\bibnamefont{Stellmer}},
  \bibinfo{author}{\bibfnamefont{P.}~\bibnamefont{Soltan-Panahi}},
  \bibinfo{author}{\bibnamefont{{\relax S. D\"orscher}}},
  \bibinfo{author}{\bibfnamefont{M.}~\bibnamefont{Baumert}},
  \bibinfo{author}{\bibfnamefont{E.~M.} \bibnamefont{Richter}},
  \bibinfo{author}{\bibnamefont{{\relax J. Kronj\"ager}}},
  \bibinfo{author}{\bibfnamefont{K.}~\bibnamefont{Bongs}}, \bibnamefont{and}
  \bibinfo{author}{\bibfnamefont{K.}~\bibnamefont{Sengstock}},
  \bibinfo{journal}{Nature Phys.} \textbf{\bibinfo{volume}{4}},
  \bibinfo{pages}{496} (\bibinfo{year}{2008}).

\bibitem[{\citenamefont{Wadkin-Snaith and Gangardt}(2012)}]{mass1}
\bibinfo{author}{\bibfnamefont{D.~C.} \bibnamefont{Wadkin-Snaith}}
  \bibnamefont{and} \bibinfo{author}{\bibfnamefont{D.~M.}
  \bibnamefont{Gangardt}}, \bibinfo{journal}{Phys. Rev. Lett}
  \textbf{\bibinfo{volume}{108}}, \bibinfo{pages}{085301}
  (\bibinfo{year}{2012}).

\bibitem[{\citenamefont{Jezek et~al.}(2016)\citenamefont{Jezek, Capuzzi, and
  Cataldo}}]{je16}
\bibinfo{author}{\bibfnamefont{D.~M.} \bibnamefont{Jezek}},
  \bibinfo{author}{\bibfnamefont{P.}~\bibnamefont{Capuzzi}}, \bibnamefont{and}
  \bibinfo{author}{\bibfnamefont{H.~M.} \bibnamefont{Cataldo}},
  \bibinfo{journal}{Phys. Rev. A} \textbf{\bibinfo{volume}{93}},
  \bibinfo{pages}{023601} (\bibinfo{year}{2016}).

\bibitem[{\citenamefont{Gallucci and Proukakis}(2016)}]{gal16}
\bibinfo{author}{\bibfnamefont{D.}~\bibnamefont{Gallucci}} \bibnamefont{and}
  \bibinfo{author}{\bibfnamefont{N.~P.} \bibnamefont{Proukakis}},
  \bibinfo{journal}{New J. Phys.} \textbf{\bibinfo{volume}{18}},
  \bibinfo{pages}{025004} (\bibinfo{year}{2016}).

\bibitem[{\citenamefont{Ryu et~al.}(2013)\citenamefont{Ryu, Blackburn, Blinova,
  and Boshier}}]{boshier13}
\bibinfo{author}{\bibfnamefont{C.}~\bibnamefont{Ryu}},
  \bibinfo{author}{\bibfnamefont{P.~W.} \bibnamefont{Blackburn}},
  \bibinfo{author}{\bibfnamefont{A.~A.} \bibnamefont{Blinova}},
  \bibnamefont{and} \bibinfo{author}{\bibfnamefont{M.~G.}
  \bibnamefont{Boshier}}, \bibinfo{journal}{Phys. Rev. Lett.}
  \textbf{\bibinfo{volume}{111}}, \bibinfo{pages}{205301}
  (\bibinfo{year}{2013}).

\bibitem[{\citenamefont{Jendrzejewski et~al.}(2014)\citenamefont{Jendrzejewski,
  Eckel, Murray, Lanier, Edwards, Lobb, and Campbell}}]{jen14}
\bibinfo{author}{\bibfnamefont{F.}~\bibnamefont{Jendrzejewski}},
  \bibinfo{author}{\bibfnamefont{S.}~\bibnamefont{Eckel}},
  \bibinfo{author}{\bibfnamefont{N.}~\bibnamefont{Murray}},
  \bibinfo{author}{\bibfnamefont{C.}~\bibnamefont{Lanier}},
  \bibinfo{author}{\bibfnamefont{M.}~\bibnamefont{Edwards}},
  \bibinfo{author}{\bibfnamefont{C.~J.} \bibnamefont{Lobb}}, \bibnamefont{and}
  \bibinfo{author}{\bibfnamefont{G.~K.} \bibnamefont{Campbell}},
  \bibinfo{journal}{Phys. Rev. Lett.} \textbf{\bibinfo{volume}{113}},
  \bibinfo{pages}{045305} (\bibinfo{year}{2014}).

\bibitem[{\citenamefont{Ryu et~al.}(2007)\citenamefont{Ryu, Andersen, Clad\'e,
  Natarajan, Helmerson, and Phillips}}]{toroidal}
\bibinfo{author}{\bibfnamefont{C.}~\bibnamefont{Ryu}},
  \bibinfo{author}{\bibfnamefont{M.~F.} \bibnamefont{Andersen}},
  \bibinfo{author}{\bibfnamefont{P.}~\bibnamefont{Clad\'e}},
  \bibinfo{author}{\bibfnamefont{V.}~\bibnamefont{Natarajan}},
  \bibinfo{author}{\bibfnamefont{K.}~\bibnamefont{Helmerson}},
  \bibnamefont{and} \bibinfo{author}{\bibfnamefont{W.~D.}
  \bibnamefont{Phillips}}, \bibinfo{journal}{Phys. Rev. Lett.}
  \textbf{\bibinfo{volume}{99}}, \bibinfo{pages}{260401}
  (\bibinfo{year}{2007}).

\bibitem[{\citenamefont{Seaman et~al.}(2007)\citenamefont{Seaman, Kr{\"{a}}mer,
  Anderson, and Holland}}]{seaman}
\bibinfo{author}{\bibfnamefont{B.~T.} \bibnamefont{Seaman}},
  \bibinfo{author}{\bibfnamefont{M.}~\bibnamefont{Kr{\"{a}}mer}},
  \bibinfo{author}{\bibfnamefont{D.~Z.} \bibnamefont{Anderson}},
  \bibnamefont{and} \bibinfo{author}{\bibfnamefont{M.~J.}
  \bibnamefont{Holland}}, \bibinfo{journal}{Phys. Rev. A}
  \textbf{\bibinfo{volume}{75}}, \bibinfo{pages}{023615}
  (\bibinfo{year}{2007}).

\bibitem[{\citenamefont{Eckel et~al.}(2014)\citenamefont{Eckel, Lee,
  Jendrzejewski, Murray, Clark, Lobb, Phillips, Edwards, and Campbell}}]{eckel}
\bibinfo{author}{\bibfnamefont{S.}~\bibnamefont{Eckel}},
  \bibinfo{author}{\bibfnamefont{J.~G.} \bibnamefont{Lee}},
  \bibinfo{author}{\bibfnamefont{F.}~\bibnamefont{Jendrzejewski}},
  \bibinfo{author}{\bibfnamefont{N.}~\bibnamefont{Murray}},
  \bibinfo{author}{\bibfnamefont{C.~W.} \bibnamefont{Clark}},
  \bibinfo{author}{\bibfnamefont{C.~J.} \bibnamefont{Lobb}},
  \bibinfo{author}{\bibfnamefont{W.~D.} \bibnamefont{Phillips}},
  \bibinfo{author}{\bibfnamefont{M.}~\bibnamefont{Edwards}}, \bibnamefont{and}
  \bibinfo{author}{\bibfnamefont{G.~K.} \bibnamefont{Campbell}},
  \bibinfo{journal}{Nature} \textbf{\bibinfo{volume}{506}},
  \bibinfo{pages}{200} (\bibinfo{year}{2014}).

\bibitem[{\citenamefont{Kumar et~al.}(2018)\citenamefont{Kumar, Dubessy, Badr,
  {\relax C. De Rossi}, {\relax M. de Go{\"{e}}r de Herve}, Longchambon, and
  Perrin}}]{perrin}
\bibinfo{author}{\bibfnamefont{A.}~\bibnamefont{Kumar}},
  \bibinfo{author}{\bibfnamefont{R.}~\bibnamefont{Dubessy}},
  \bibinfo{author}{\bibfnamefont{T.}~\bibnamefont{Badr}},
  \bibinfo{author}{\bibnamefont{{\relax C. De Rossi}}},
  \bibinfo{author}{\bibnamefont{{\relax M. de Go{\"{e}}r de Herve}}},
  \bibinfo{author}{\bibfnamefont{L.}~\bibnamefont{Longchambon}},
  \bibnamefont{and} \bibinfo{author}{\bibfnamefont{H.}~\bibnamefont{Perrin}},
  \bibinfo{journal}{Phys. Rev. A} \textbf{\bibinfo{volume}{97}},
  \bibinfo{pages}{043615} (\bibinfo{year}{2018}).

\bibitem[{\citenamefont{Murray et~al.}(2013)\citenamefont{Murray, Krygier,
  Edwards, Wright, Campbell, and Clark}}]{mu13}
\bibinfo{author}{\bibfnamefont{N.}~\bibnamefont{Murray}},
  \bibinfo{author}{\bibfnamefont{M.}~\bibnamefont{Krygier}},
  \bibinfo{author}{\bibfnamefont{M.}~\bibnamefont{Edwards}},
  \bibinfo{author}{\bibfnamefont{K.~C.} \bibnamefont{Wright}},
  \bibinfo{author}{\bibfnamefont{G.~K.} \bibnamefont{Campbell}},
  \bibnamefont{and} \bibinfo{author}{\bibfnamefont{C.~W.} \bibnamefont{Clark}},
  \bibinfo{journal}{Phys. Rev. A} \textbf{\bibinfo{volume}{88}},
  \bibinfo{pages}{053615} (\bibinfo{year}{2013}).

\bibitem[{\citenamefont{Castin and Dum}(1999)}]{castin}
\bibinfo{author}{\bibfnamefont{Y.}~\bibnamefont{Castin}} \bibnamefont{and}
  \bibinfo{author}{\bibfnamefont{R.}~\bibnamefont{Dum}}, \bibinfo{journal}{Eur.
  Phys. J. D} \textbf{\bibinfo{volume}{7}}, \bibinfo{pages}{399}
  (\bibinfo{year}{1999}).

\bibitem[{\citenamefont{Carr et~al.}(2000)\citenamefont{Carr, Clark, and
  Reinhardt}}]{carr}
\bibinfo{author}{\bibfnamefont{L.~D.} \bibnamefont{Carr}},
  \bibinfo{author}{\bibfnamefont{C.~W.} \bibnamefont{Clark}}, \bibnamefont{and}
  \bibinfo{author}{\bibfnamefont{W.~P.} \bibnamefont{Reinhardt}},
  \bibinfo{journal}{Phys. Rev. A} \textbf{\bibinfo{volume}{62}},
  \bibinfo{pages}{063610} (\bibinfo{year}{2000}).

\bibitem[{\citenamefont{Nguyen et~al.}(2014)\citenamefont{Nguyen, Dyke, Luo,
  Malomed, and Hulet}}]{ngu14}
\bibinfo{author}{\bibfnamefont{J.~H.~V.} \bibnamefont{Nguyen}},
  \bibinfo{author}{\bibfnamefont{P.}~\bibnamefont{Dyke}},
  \bibinfo{author}{\bibfnamefont{D.}~\bibnamefont{Luo}},
  \bibinfo{author}{\bibfnamefont{B.~A.} \bibnamefont{Malomed}},
  \bibnamefont{and} \bibinfo{author}{\bibfnamefont{R.~G.} \bibnamefont{Hulet}},
  \bibinfo{journal}{Nature Phys.} \textbf{\bibinfo{volume}{10}},
  \bibinfo{pages}{918} (\bibinfo{year}{2014}).

\bibitem[{\citenamefont{Stellmer et~al.}(2008)\citenamefont{Stellmer, Becker,
  Soltan-Panahi, Richter, D{\"{o}}rscher, Baumert, Kronj{\"{a}}ger, Bongs, and
  Sengstock}}]{stellmer08}
\bibinfo{author}{\bibfnamefont{S.}~\bibnamefont{Stellmer}},
  \bibinfo{author}{\bibfnamefont{C.}~\bibnamefont{Becker}},
  \bibinfo{author}{\bibfnamefont{P.}~\bibnamefont{Soltan-Panahi}},
  \bibinfo{author}{\bibfnamefont{E.-M.} \bibnamefont{Richter}},
  \bibinfo{author}{\bibfnamefont{S.}~\bibnamefont{D{\"{o}}rscher}},
  \bibinfo{author}{\bibfnamefont{M.}~\bibnamefont{Baumert}},
  \bibinfo{author}{\bibfnamefont{J.}~\bibnamefont{Kronj{\"{a}}ger}},
  \bibinfo{author}{\bibfnamefont{K.}~\bibnamefont{Bongs}}, \bibnamefont{and}
  \bibinfo{author}{\bibfnamefont{K.}~\bibnamefont{Sengstock}},
  \bibinfo{journal}{Phys. Rev. Lett.} \textbf{\bibinfo{volume}{101}},
  \bibinfo{pages}{120406} (\bibinfo{year}{2008}).

\bibitem[{\citenamefont{Huang et~al.}(2001)\citenamefont{Huang, Velarde, and
  Makarov}}]{collision}
\bibinfo{author}{\bibfnamefont{G.}~\bibnamefont{Huang}},
  \bibinfo{author}{\bibfnamefont{M.~G.} \bibnamefont{Velarde}},
  \bibnamefont{and} \bibinfo{author}{\bibfnamefont{V.~A.}
  \bibnamefont{Makarov}}, \bibinfo{journal}{Phys. Rev. A}
  \textbf{\bibinfo{volume}{64}}, \bibinfo{pages}{013617}
  (\bibinfo{year}{2001}).

\bibitem[{\citenamefont{Kavoulakis and Pethick}(1998)}]{kavoulakis98}
\bibinfo{author}{\bibfnamefont{G.~M.} \bibnamefont{Kavoulakis}}
  \bibnamefont{and} \bibinfo{author}{\bibfnamefont{C.~J.}
  \bibnamefont{Pethick}}, \bibinfo{journal}{Phys. Rev. A}
  \textbf{\bibinfo{volume}{58}}, \bibinfo{pages}{1563} (\bibinfo{year}{1998}).

\end{thebibliography}
\providecommand{\noopsort}[1]{}\providecommand{\singleletter}[1]{#1}%

\end{document}